\documentclass[conference]{IEEEtran}
\usepackage{graphicx}
\usepackage{cite}
\usepackage{hyperref}
\usepackage{amsmath}
\usepackage{amsfonts}
\title{Substituting Proof of Work in Blockchain with Training-Verified Collaborative Model Computation}

\author{
\IEEEauthorblockN{Mohammad Ishzaz Asif Rafid}

\and
\IEEEauthorblockN{Morsalin Sakib}

}

\begin{document}

\maketitle
\begin{abstract}

Bitcoin's Proof of Work (PoW) mechanism, while central to achieving decentralized consensus, has been widely criticized for its excessive energy consumption and hardware inefficiencies \cite{devries2018bitcoin, truby2018decarbonizing}. This paper proposes a hybrid architecture that replaces Bitcoin’s traditional PoW with a centralized, cloud-based collaborative training system. In this model, miners contribute their computational resources to train segments of horizontally scaled machine learning models on preprocessed datasets, ensuring data privacy and producing meaningful computational outputs \cite{li2017securing}. A centralized server evaluates contributions based on two metrics: the number of parameters successfully trained and the reduction in model loss achieved during a fixed training cycle. At the end of each cycle, a weighted lottery determines the winning miner, who receives a digitally signed certificate from the server. This certificate serves as a verifiable substitute for PoW and grants the right to append a new block to the blockchain \cite{nakamoto2008bitcoin}. By integrating cryptographic digital signatures and SHA-256 hashing \cite{nist2015sha}, this approach maintains blockchain integrity while transforming the energy-intensive mining process into one that provides real-world computational value.
\end{abstract}

\section{Introduction}

Since its launch in 2008, Bitcoin has used Proof of Work (PoW) as a consensus algorithm to validate transactions and secure the network \cite{devries2018bitcoin}. PoW entails that miners solve computationally demanding cryptographic puzzles, which uses substantial amounts of energy and hardware resources \cite{truby2018decarbonizing}. As the network of Bitcoin has grown, so has its carbon footprint, which has drawn the ire of environmentalists, policymakers, and computer scientists alike \cite{truby2018decarbonizing, gervais2016security}.
This paper solves that fundamental inefficiency by presenting an innovative solution: substituting Bitcoin's conventional PoW with a centralized server that provides a cryptographically verifiable digital signature when legitimate real-world computational work is verified. Instead of expending computational energy on meaningless hashing, participants contribute general-purpose computing resources via a cloud platform. A centralized authority, which has kernel-level access to the contributor's machine, guarantees that only legitimate computational work is compensated \cite{li2017securing}. Upon verification, the server generates a signed "Proof of Work" document, which miners can utilize to insert blocks onto the blockchain.

This model keeps the same block structure, consensus integrity, and SHA-256 hash use \cite{nist2015sha}, but outsources effort from bare hash calculations to beneficial cloud work. While this adds a degree of centralization, the efficiency and sustainability benefits are profound. This paper details how Bitcoin's PoW presently works, the inner workings of SHA-256 and digital signatures, and how the proposed system might provide a secure, functional, and more environmentally friendly alternative.

\section{Literature Review}

Bitcoin's Proof of Work (PoW), first introduced by Nakamoto \cite{nakamoto2008bitcoin}, remains the most widely adopted consensus mechanism in blockchain systems. While its cryptoeconomic security is robust, numerous studies have underscored its inefficiencies. De Vries \cite{devries2018bitcoin} estimated Bitcoin’s annualized energy consumption to rival that of mid-sized nations, while Truby \cite{truby2018decarbonizing} emphasized the unsustainable environmental implications of this energy usage. Gervais et al. \cite{gervais2016security} further analyzed how PoW’s resource-intensive nature constrains blockchain scalability.

To address these issues, multiple alternatives have been proposed. Proof of Stake (PoS) mechanisms, as pioneered in Peercoin \cite{king2012peercoin}, substitute computational work with stake-based validation, drastically reducing energy demands. Other mechanisms, such as Proof of Space \cite{ateniese2014proofs} and Proof of Activity \cite{bentov2014proof}, attempt to optimize resource utilization. However, these approaches introduce trade-offs in terms of decentralization, fairness, or susceptibility to long-range attacks.

Recent work has explored integrating useful computation into consensus protocols. Eyal and Sirer \cite{eyal2014majority} examined centralization dynamics in mining pools, while Li et al. \cite{li2017securing} investigated hybrid consensus mechanisms involving off-chain computations. Parallel research in distributed deep learning \cite{ma2018sgd, hedge2015parallel, su2015modelavg} has advanced scalable, horizontally partitioned model training using stochastic gradient descent (SGD) and asynchronous updates. These methods leverage GPUs and large compute clusters to achieve near-linear training speedups \cite{paine2013gpu, najafabadi2019hpcc}, making them well-suited for resource-constrained consensus applications.

Kaplan et al. \cite{kaplan2020scaling} established scaling laws for large language models, providing a theoretical basis for horizontally scaling deep neural networks. Chaubard et al. \cite{chaubard2024gaf} introduced Gradient Agreement Filtering to improve distributed optimization in multi-participant environments. These insights inform the proposed system, which reframes PoW as Proof of Useful Training, aligning blockchain consensus with productive machine learning tasks.

\section{Background}

\subsection{Bitcoin Blockchain Architecture}
Bitcoin's blockchain is a distributed ledger, with each block having:
A block header (version, timestamp, previous block hash, Merkle root).

A nonce utilized for computing PoW.

A collection of verified transactions.

Every new block includes a reference to the previous block's hash, forming a tamper-evident chain. Miners race to discover a nonce that, when hashed together with the block information by SHA‑256, results in a value less than a dynamic target difficulty \cite{nist2015sha}. Network security is established through this computational effort \cite{devries2018bitcoin}.

\begin{figure}[h]

\centering
\includegraphics[width=\linewidth]{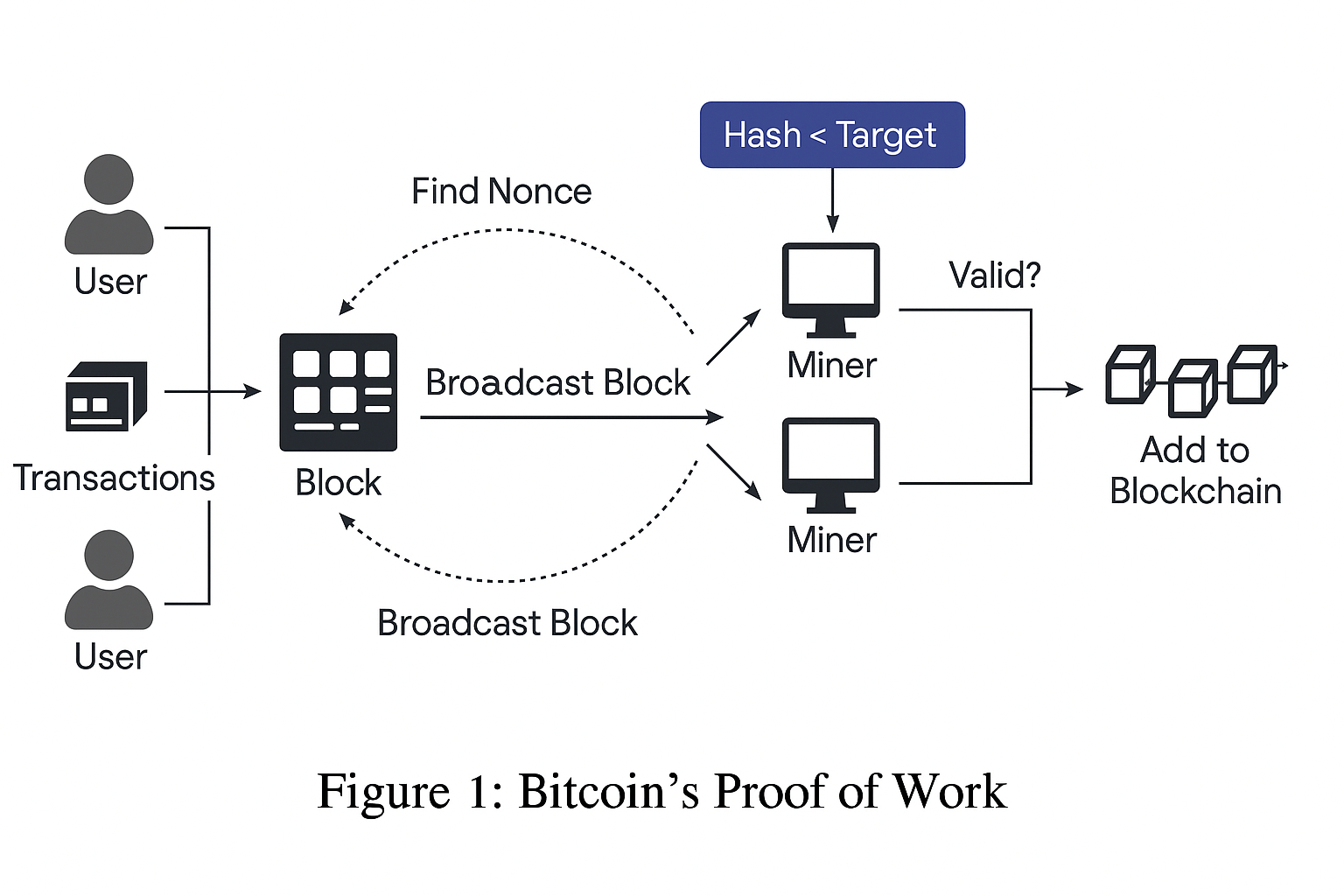}
\caption{Traditional Bitcoin Proof of Work Workflow.}
\label{fig:bitcoin-pow}
\end{figure}
\subsection{Proof of Work Mechanism}

PoW makes miners conduct a brute-force search for a valid nonce. This process does not accomplish inherently useful work—it is only done to secure the network by making the creation of blocks expensive. Miners who succeed are compensated in newly created bitcoins and transaction fees. Although this scheme precludes Sybil attacks and adversarial rewriting of history, it leads to huge energy usage \cite{truby2018decarbonizing}.
\subsection{SHA-256: Function and Purpose}

The SHA‑256 (Secure Hash Algorithm 256-bit) is a 256-bit cryptographic hash function standardized by the NSA \cite{nist2015sha}. It takes input data of any length and produces a fixed 256-bit hash. Its most important properties are:
Preimage resistance: It is impossible to reverse-engineer input from the hash.

Collision resistance: Very improbable to have two distinct inputs with the same hash.

Avalanche effect: Small changes to the inputs radically transform the hash.

SHA‑256 is used twice (double SHA‑256) in Bitcoin to provide additional resistance to length extension attacks. This hashing forms the basis of block linking and transaction confirmation.

\subsection{Digital Signatures and ECDSA}

Bitcoin employs the Elliptic Curve Digital Signature Algorithm (ECDSA) to confirm ownership of funds. A digital signature is obtained by pairing a private key with a cryptographic function applied to data related to transactions. This allows for:
Authentication: The signature can be generated by the private key holder alone.

Non-repudiation: The signer cannot reject their participation.

Integrity: Any tampering invalidates the signature.

Mathematically, ECDSA is about creating a random integer

k, computing a curve point
(x,y), and yielding a signature pair

(r,s) depending on the message hash and private key. This makes the signature verifiable by anyone who has the relevant public key.

\section{Proposed System}

\subsection{System Overview}
The proposed system transforms consensus into a collaborative model training task, redefining Proof of Work as Proof of Useful Training. Instead of competing in hash-based nonce races, miners allocate GPU/CPU resources to train segments of horizontally scaled neural networks. Models are partitioned across layers or parameter shards to enable parallel execution across multiple nodes, following principles from large-scale distributed training \cite{ma2018sgd, kaplan2020scaling, li2021deploy}.

Each training epoch lasts 20 minutes, during which miners:

Receive preprocessed dataset partitions and model segments (ensuring privacy and preventing raw data reconstruction).

Train their allocated segments using stochastic gradient descent (SGD) with GPU acceleration \cite{paine2013gpu}.

Upload updated weights to the coordination server.

The server evaluates contributions using two metrics:

Parameter volume: Quantifying the scale of updates (reflecting computational effort).

Loss reduction: Measuring training effectiveness.

These metrics determine each miner’s lottery weight. A weighted, auditable lottery then selects a block producer. The winner receives a digitally signed certificate, serving as a verifiable proof of useful work, granting the right to propose a block.

\subsection{Proof of Work via Model Training}
This system redefines ``Proof of Work'' as ``Proof of Useful Training.'' Instead of computing meaningless hashes, miners allocate their resources to train neural network segments using distributed stochastic gradient descent (SGD). Their contributions are assessed based on two key factors: the number of model parameters they successfully update and the amount of loss reduction achieved during the training cycle. These values are combined into a single contribution score that determines each miner's probability of being selected in a weighted lottery. The miner with the highest weighted chance, chosen via this lottery, earns the right to produce the next block. 

\begin{figure}[htbp]
    \centering
    \includegraphics[width=0.9\linewidth]{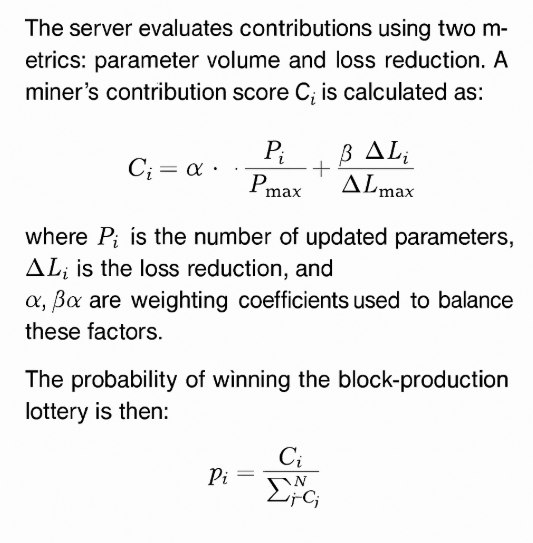}
    \caption{Miner contribution scoring and weighted lottery selection.}
    \label{fig:contribution-formula}
\end{figure}

\subsection{Digital Signature as Proof of Work}
After the server completes the lottery, it provides a digitally signed certificate to the winning miner. This certificate is the new PoW and contains:
The miner’s public key.

Resource contribution measures.

Cycle timestamp and unique nonce.

Server's digital signature (ECDSA).

The successful miner incorporates this certificate into the block header. As the block spreads across the network, any node can verify its genuineness with the server's public key, making the block production process transparent and trustworthy.

This certificate serves as the novel "proof of work." Its authenticity can be verified by any node on the network using the public key of the server. This maintains the traditional PoW verifiability without the need for decentralized brute-force mining.

\section{System Workflow}

\subsection{Block Creation Process}

Registration: Miners register for the upcoming 20-minute training cycle.

Model and Data Distribution: The server provides miners with a preprocessed dataset and a horizontally scaled model segment to train.

Model Training: Miners train their assigned layers using their available GPU/CPU resources for the 20-minute window.

Contribution Evaluation: Miners upload updated weights. The server evaluates each miner’s contribution based on parameters trained and loss reduction achieved.

Lottery Execution: A weighted lottery selects the block producer, giving higher chances to miners with greater contributions.

Certificate Issuance: All miners receive signed contribution certificates; the winner receives an exclusive block-signing certificate.

Block Assembly and Broadcasting: The winning miner creates and broadcasts the block, embedding the certificate in the block header.

Network Validation: Other nodes verify the block’s certificate using the server’s public key.

Reward Distribution: The winning miner receives the block reward.

This workflow transforms blockchain mining into a process that produces tangible outcomes — trained machine learning models — while maintaining fair block selection and cryptographic verifiability.

\begin{figure}[h]

\centering
\includegraphics[width=\linewidth]{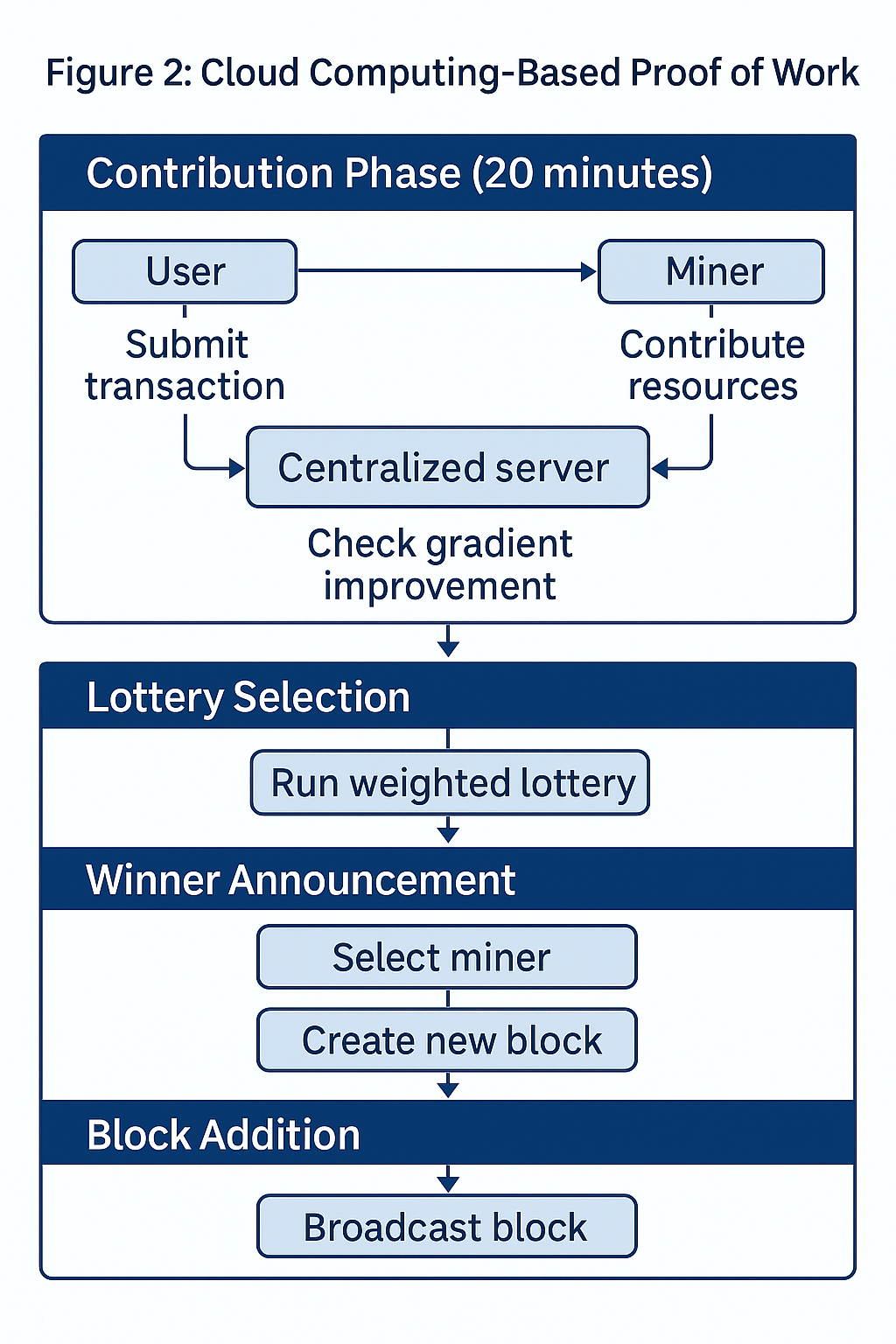}
\caption{Traditional Bitcoin Proof of Work Workflow.}
\label{fig:bitcoin-pow}
\end{figure}
\begin{figure}[h]

\centering
\includegraphics[width=\linewidth]{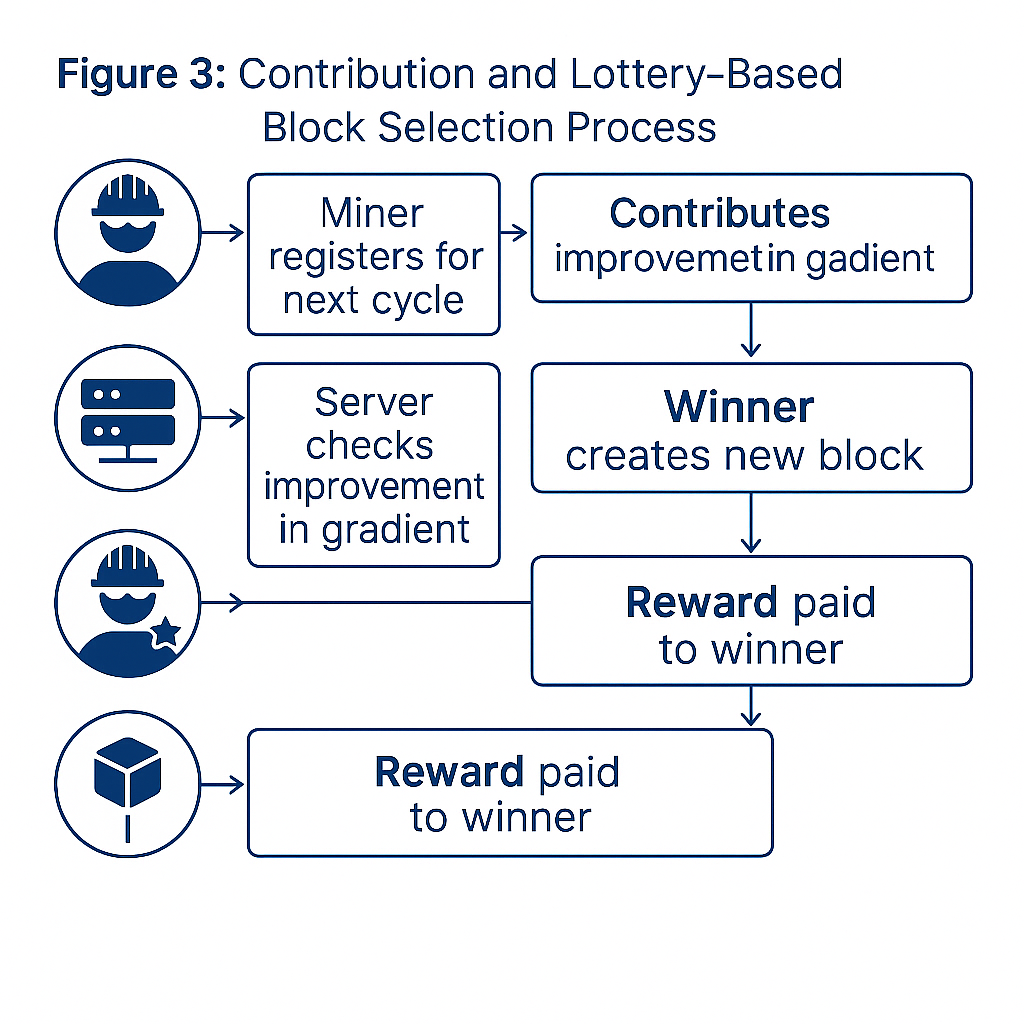}
\caption{Traditional Bitcoin Proof of Work Workflow.}
\label{fig:bitcoin-pow}
\end{figure}
\begin{figure}[h]

\centering
\includegraphics[width=\linewidth]{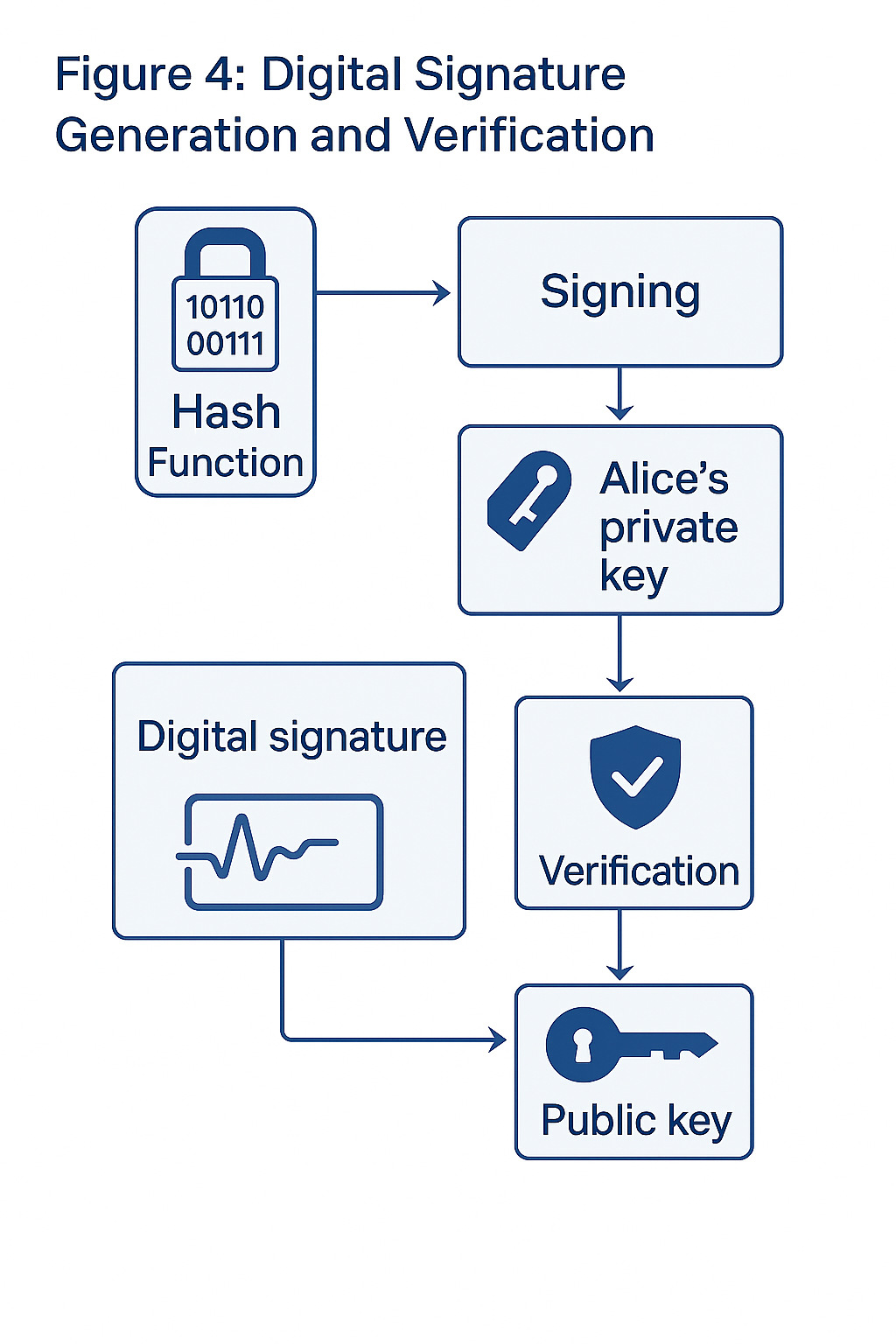}
\caption{Traditional Bitcoin Proof of Work Workflow.}
\label{fig:bitcoin-pow}
\end{figure}
\section{Comparative Analysis}

\subsection{Energy Efficiency}

Bitcoin’s PoW network consumes approximately 129 TWh annually \cite{truby2018decarbonizing}, with 100\% of this energy directed toward cryptographic hashing — a process with no real-world utility. In contrast, the proposed training-based model recycles this energy into meaningful computations, training horizontally scaled models that can support enterprise, scientific, or AI applications \cite{li2021deploy, prasad2025scalability}. Distributed training frameworks like Horovod and DeepSpeed have demonstrated that large-scale models can be trained efficiently across heterogeneous clusters \cite{hedge2015parallel, najafabadi2019hpcc}, enabling this consensus approach to deliver dual benefits: network security and AI model development.

\begin{figure}[h]

\centering

\includegraphics[width=\linewidth]{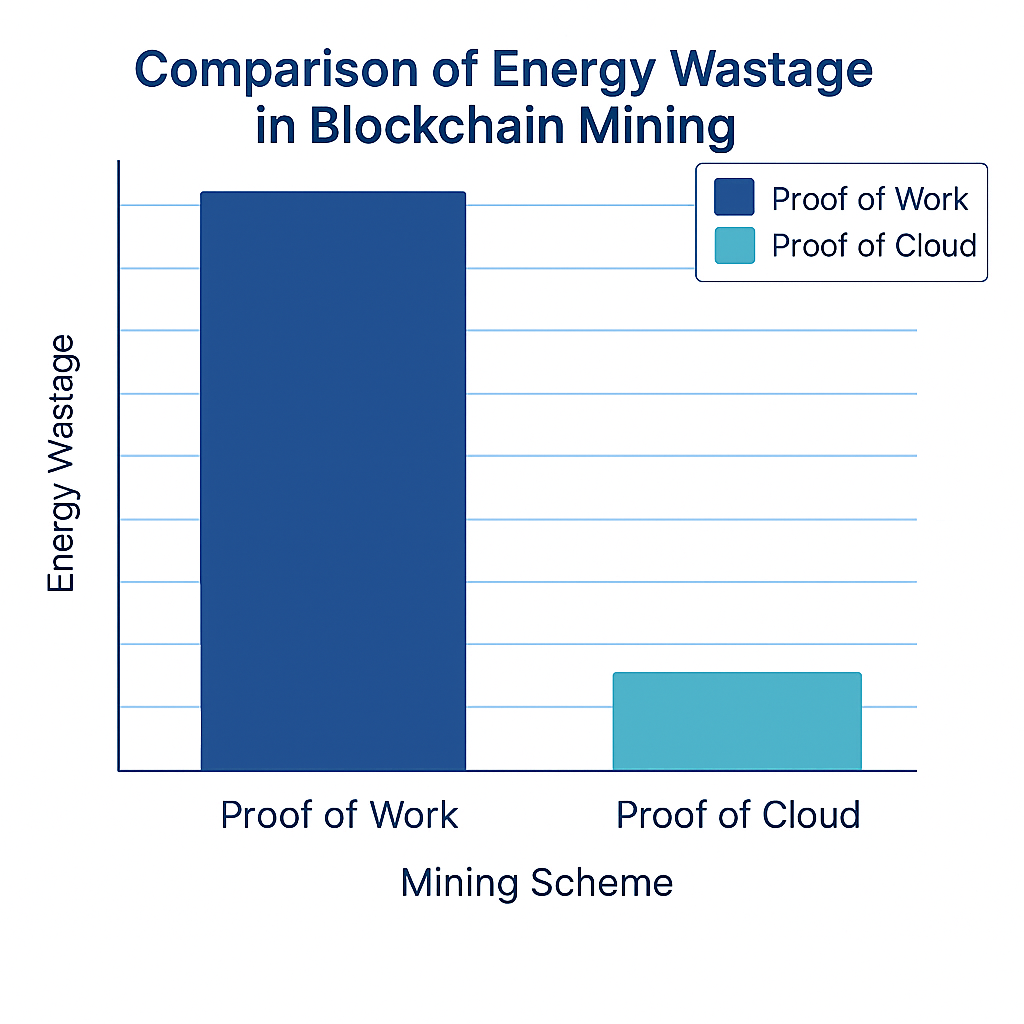}

\caption{Energy Consumption Comparison: Bitcoin and Proposed Model.}

\label{fig:energy-comparison}

\end{figure}

\subsection{Security Considerations}
By using a weighted lottery based on model training contributions, this mechanism blends elements of Proof of Stake and verifiable computing. Digitally signed certificates (ECDSA) provide cryptographic integrity, while commit-reveal schemes or zero-knowledge proofs can make contribution scoring auditable \cite{chaubard2024gaf}. This enhances transparency and mitigates risks of manipulation by the centralized task server.

\subsection{Centralization Trade-offs}
The introduction of a coordinating server mirrors concerns raised by Eyal and Sirer \cite{eyal2014majority} regarding centralization in mining pools. To address this, federated server architectures and verifiable distributed aggregation could be implemented, reducing single-point-of-failure risks while maintaining operational efficiency.

\section{Challenges and Limitations}

\subsection{Trust in the Centralized Server}
The proposed framework introduces a centralized coordination server responsible for task assignment, contribution evaluation, and certificate issuance. While digitally signed certificates ensure verifiability, participants must trust that the server assigns tasks fairly and reports contributions honestly. This dependence creates potential vulnerabilities if the server acts maliciously or colludes with participants.

Research in verifiable computation offers potential mitigations. Zero-knowledge proofs (ZKPs) have been proposed to verify off-chain computations without revealing underlying data or requiring full recomputation \cite{chaubard2024gaf}. Similarly, verifiable outsourcing frameworks such as those explored by Ateniese et al. \cite{ateniese2014proofs} could provide cryptographic guarantees that training contributions are authentic. Incorporating these mechanisms would enhance transparency and reduce reliance on server trustworthiness.

\subsection{Single Point of Failure}
A centralized server also introduces a single point of failure. Any downtime, targeted attack, or system compromise could halt block production. This risk parallels concerns highlighted by Eyal and Sirer \cite{eyal2014majority} about centralization in mining pools. To mitigate this, the architecture could evolve into a federated consensus network, where multiple independently operated servers share coordination duties, using Byzantine Fault Tolerant (BFT) protocols for agreement. Similar federated models have been successfully deployed in distributed learning and high-performance computing clusters \cite{hedge2015parallel, prasad2025scalability}, suggesting that a multi-server approach could significantly improve system resilience.

\subsection{Regulatory Implications}
By introducing a coordinating entity, the proposed model becomes more visible to regulatory oversight. This could result in constraints on its deployment in permissionless public networks, echoing concerns discussed in blockchain governance literature \cite{truby2018decarbonizing}. However, for enterprise or research-focused blockchains, where a degree of trust in authorities is acceptable, such oversight could be beneficial, ensuring compliance with legal and ethical frameworks for data use and computation.

\subsection{Scalability and Performance Bottlenecks}
Horizontally scaled model training inherently improves parallelism \cite{kaplan2020scaling, li2021deploy}, but the centralized server’s evaluation and aggregation process could become a bottleneck as participant numbers grow. Previous studies on distributed SGD highlight that communication overhead often dominates as the system scales \cite{ma2018sgd, paine2013gpu}. Techniques such as gradient compression, decentralized aggregation, and hierarchical parameter servers could alleviate these bottlenecks, enabling the system to maintain high throughput even with thousands of participants.

Incorporating these scaling techniques — along with dynamic task allocation based on participant hardware profiles — could improve efficiency while maintaining fairness across heterogeneous miner capabilities.

\section{Conclusion and Future Work}

This paper introduced a novel consensus mechanism that substitutes Bitcoin's energy-intensive Proof of Work (PoW) with a \textit{training-based Proof of Work} model. In this approach, miners repurpose their computational resources to train segments of horizontally scaled machine learning models using preprocessed datasets, ensuring data privacy and generating real-world value. The server evaluates miners' contributions based on the number of parameters trained and the reduction in model loss, subsequently conducting a weighted lottery to select the block producer. The selected miner receives a digitally signed certificate, which serves as verifiable proof of meaningful work and enables block creation.  

This model significantly improves \textbf{energy efficiency} by redirecting mining efforts to beneficial computations, while maintaining \textbf{cryptographic verifiability} and \textbf{blockchain integrity}. Although the inclusion of a centralized server introduces trade-offs in terms of decentralization and creates a potential single point of failure, these challenges can be mitigated through federated server networks and the use of transparency-enhancing cryptographic techniques such as zero-knowledge proofs.  

Future research will focus on developing federated or decentralized server architectures, implementing privacy-preserving verification mechanisms, and benchmarking the system's energy efficiency and model training performance compared to traditional PoW networks. This work opens a pathway for integrating blockchain consensus mechanisms with real-world AI and machine learning tasks, transforming block production into a process that is both sustainable and socially beneficial.

\bibliographystyle{IEEEtran} 
\bibliography{references}
\end{document}